\documentclass[aip,graphicx]{revtex4-1}

\usepackage{amssymb}
\usepackage{graphicx}
\usepackage{amsmath}
\usepackage{amsfonts}
\usepackage{epsfig}
\usepackage{epstopdf}

\draft

\begin{document}

\title{ \textbf{Theoretical Model of Superconducting Spintronic SIsFS Devices }}
\author{S.~V.~Bakurskiy}
\affiliation{Lomonosov Moscow State University Skobeltsyn Institute of Nuclear Physics
(MSU SINP), Leninskie gory, GSP-1, Moscow 119991, Russian Federation }
\affiliation{Faculty of Physics, Lomonosov Moscow State University, 119992 Leninskie
Gory, Moscow, Russian Federation}
\author{N.~V.~Klenov}
\affiliation{Lomonosov Moscow State University Skobeltsyn Institute of Nuclear Physics
(MSU SINP), Leninskie gory, GSP-1, Moscow 119991, Russian Federation }
\affiliation{Faculty of Physics, Lomonosov Moscow State University, 119992 Leninskie
Gory, Moscow, Russian Federation}
\author{I.~I.~Soloviev}
\affiliation{Lomonosov Moscow State University Skobeltsyn Institute of Nuclear Physics
(MSU SINP), Leninskie gory, GSP-1, Moscow 119991, Russian Federation }
\author{V.~V.~Bol'ginov}
\affiliation{Institute of Solid State Physics, Russian Academy of Sciences,
Chernogolovka, 142432, Russian Federation}
\author{V.~V.~Ryazanov}
\affiliation{Institute of Solid State Physics, Russian Academy of Sciences,
Chernogolovka, 142432, Russian Federation}
\author{I.~V.~Vernik}
\affiliation{HYPRES, Inc. 175 Clearbrook Rd., Elmsford, NY 10523 USA}
\author{O.~A.~Mukhanov}
\affiliation{HYPRES, Inc. 175 Clearbrook Rd., Elmsford, NY 10523 USA}
\author{M.~Yu.~Kupriyanov}
\affiliation{Lomonosov Moscow State University Skobeltsyn Institute of Nuclear Physics
(MSU SINP), Leninskie gory, GSP-1, Moscow 119991, Russian Federation }
\author{A.~A.~Golubov}
\affiliation{Faculty of Science and Technology and MESA+ Institute for Nanotechnology,
University of Twente, 7500 AE Enschede, The Netherlands}
\date{\today }
\date{\today }

\begin{abstract}
Motivated by recent progress in development of cryogenic memory compatible with single flux quantum (SFQ) circuits we have performed a theoretical study of magnetic SIsFS Josephson junctions, where 'S' is a bulk superconductor, 's' is a thin superconducting film, 'F' is a metallic ferromagnet and 'I' is an insulator. We  calculate the Josephson current as a function of  s and F layers thickness, temperature and exchange energy of F film. We outline several modes of operation of these junctions and demonstrate their unique ability to have large product of a critical current $I_{C}$ and a normal-state resistance $R_{N}$ in the $\pi$ state, comparable to that in SIS tunnel junctions commonly used in SFQ circuits. We develop a model describing switching of the Josephson critical current in these devices by external magnetic field. The results are in good agreement with the experimental data for Nb-Al/AlO${_x}$-Nb-Pd$_{0.99}$Fe$_{0.01}$-Nb junctions.
\end{abstract}

\maketitle

\bigskip

Practical applications of superconducting digital circuits were significantly limited by the relatively low capacity of superconducting memory. This motivated initial proposals to use superconductor/ferromagnet (S/F) hybrid structures as basis for the development in cryogenic magnetic Random Access Memories (RAMs) \cite{Oh,Tagirov,Held}.
Following the first experimental realization
of SFS Josephson junctions \cite{RyazanovU,Ryazanov}, much attention was paid to realize  Josephson devices with complex magnetic barriers
allowing switching between high and low critical currents.
A number of different device structures were considered
\cite{VolkovAF,GKF1,VolkovAF2,Houzet,Karminskaya1,Karminskaya2,Gabor,BirgeLR, Bolginov}
based either on superconducting spintronics effects or singlet-triplet switching within the magnetic barrier.
However, these approaches were based on structures with reduced characteristic voltage $I_cR_N$ of junctions.

Recently, successful realization of switchable Nb-Al/AlO${_x}$-Nb-Pd$_{0.99}$Fe$_{0.01}$-Nb junctions was reported in \cite{Larkin, Vernik, Ryazanov3}.
These junctions are of SIsFS type, i.e., a serial connection of the SIs tunnel junction and sFS sandwich.
SIsFS structure has high characteristic voltage, $I_cR_N,$ due to the presence of tunnel barrier 'I'.
At the same time the whole structure behaves as a single junction with respect to an external magnetic field $H_{ext}$
and magnetic flux $\Phi$ penetrating into the structure, since the intermediate layer s is too thin to screen magnetic field.
As a result, the magnetic field entering the Pd$_{0.99}$Fe$_{0.01}$ layer will modify its effective magnetization,
facilitating the critical current control of the whole double-barrier SIsFS structure.
According to Ref. \cite{Uspenskaya}, effective magnetization in the dilute Pd$_{0.99}$Fe$_{0.01}$
is controlled by Fe-rich Pd$_3$Fe nanoclusters, which can be easily reordered by a weak magnetic field.
\begin{figure}[tbh]
\begin{center}
\includegraphics[width=5.5cm]{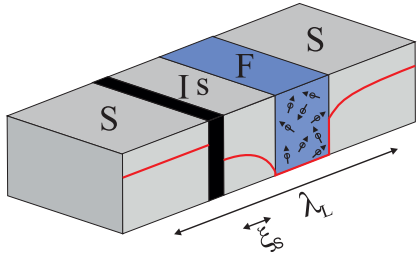}
\end{center}
\caption{Schematic design of SIsFS Josephson junction.
Solid line demonstrates typical distribution of pair
potential $\Delta $ along the structure. It reaches bulk values in the S
electrodes, is suppressed in the middle s layer and vanishes in the ferromagnetic
region F. The characteristic length scales are also marked in the figure:
$\protect\lambda _{L}$ is the London penetration depth and $\protect\xi_{S} $
is the coherence length typical for niobium.}
\end{figure}
The purpose of this paper is to develop a theory describing various modes of operation of SIsFS Josephson devices.
We propose a quasiclassical model for a double-barrier Josephson structure in which magnetic and spin states of the F-film in an sFS part of the structure can change the properties and even the ground state of its Sis part.
As a result, an external magnetic field may switch the junction from a superconducting to a resistive state or may transform a conventional current-phase relation to an inverted one.
We compare the results with experimental data recently obtained for for Nb-Al/AlO${_x}$-Nb-Pd$_{0.99}$Fe$_{0.01}$-Nb junctions.

We consider the multilayered structure presented in Fig.\textbf{1}. 
It consists of two superconducting electrodes 'S' separated by a tunnel barrier 'I',
an intermediate thin superconducting film 's' and a ferromagnetic layer 'F'.
To describe the supercurrent transport in the structure, we assume that the conditions of
a dirty limit are fulfilled for all metals. We also assume that all superconducting films in the structure
are made from identical materials, i.e., they can
be described by the same critical temperature, $T_{C},$ and coherence length, $%
\xi _{S}=(D_{S}/2\pi T_{C})^{1/2}$, where $D_{S}$ is the electronic diffusion
coefficient. The tunnel barrier I and the sF and FS interfaces are characterized, respectively, by
the following parameters
$\gamma_{BI}=R_{I}\mathcal{A}/\xi_{S}\rho_{S},$ $\gamma_{BFS}=R_{FS}\mathcal{A}/\xi_{F}\rho_{F},$
and $\gamma=\rho_{S}\xi_{S}/\xi_{F}\rho_{F}.$
Here $\mathcal{A},$ $R_{I}$ and $R_{FS}$ are the area and the resistances of the interfaces, $\xi_{S,F}$ and $\rho_{S,F}$
are the coherence lengths and normal state resistivities of S and F materials, respectively.
Under the above conditions the Josephson effect in the SIsFS junctions can be described by solving the Usadel
equations \cite{Usadel, RevG, RevB, RevV} with Kupriyanov-Lukichev (KL) boundary conditions \cite%
{KL} at Is, sF and FS interfaces and with the bulk pair potential in the depth of S-electrodes.

\begin{figure}[tbh]
\begin{center}
\includegraphics[width=8.5cm]{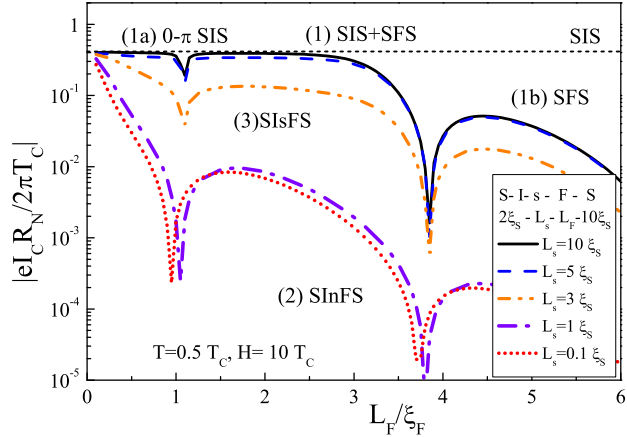}
\end{center}
\caption{Characteristic voltage $I_{C}R_{N}$ of the SIsFS structures
versus thickness of the F-layer $L_{F}$ for different thicknesses of the middle
superconducting film $L_{s}$ at $T=0.5T_{C}$. Short-dashed straight line shows
the $I_{C}R_{N}$ product of the tunnel SIS junction at the same temperature.
Interface parameters: $\gamma _{BI}=1000$, $\gamma_{BFS}=0.3$ and $\gamma=1$ at the sF and FS interfaces.}
\label{ICRN_LF}
\end{figure}

\begin{figure}[tbh]
\begin{center}
\includegraphics[width=8.5cm]{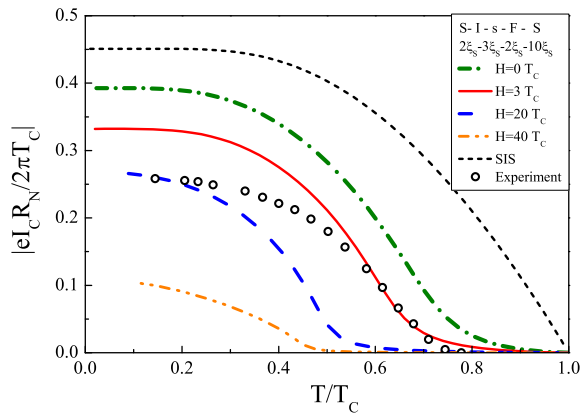}
\end{center}
\caption{The temperature dependence of characteristic voltage $I_{C}R_{N}$
of SIsFS structure for different values of exchange field $H$ in the F-layer.
The short-dashed line demonstrates the dependence characteristic for a conventional tunnel
SIS Junction. It is seen how the exchange field H shifts the
effective critical temperature $T_{C}^{\ast }$, corresponding to the switching
of the s-layer from the superconducting state to the normal one. The circles show
$I_{C}R_{N}$ measured in Nb-Al/AlO${_x}$-Nb-Pd$_{0.99}$Fe$_{0.01}$-Nb junctions \protect\cite{Vernik},
proving the existence of effective critical temperature $T_{C}^{\ast }$ in these samples.}
\label{Temp1}
\end{figure}

\begin{figure}[tbh]
\begin{center}
\includegraphics[width=8.5cm]{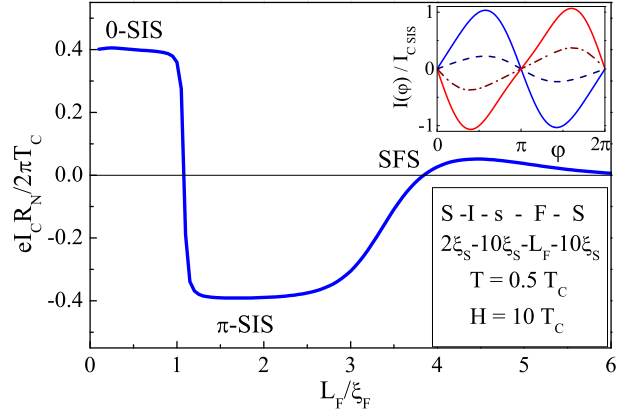}
\end{center}
\caption{The dependence of characteristic voltage $I_{C}R_{N}$ on
F-layer thickness $L_{F}$ in the SIsFS structure with the s-layer in
the superconducting state. Inset shows the current-phase relation in the
vicinity of the first $0-\protect\pi $ transition. Switching
from $0$ to $\protect\pi $ state in the mode (1a) preserves the value of
critical current $I_{C}$ as well as characteristic voltage $I_{C}R_{N}$.}
\label{ICRN_FIG}
\end{figure}

The formulated above boundary problem has been solved numerically.
The results are summarized in Figs.2-4, where various modes of operation of the structure are defined
according to chosen materials and layer thicknesses. These modes are
clearly defined by the dependencies of characteristic voltage $I_{C}R_{N}$
on thickness of intermediate superconductor, $L_{s},$ and ferromagnetic, $L_{F},$ layers %
(see Fig.\ref{ICRN_LF} - Fig.\ref{ICRN_FIG}).

\emph{Mode (1) in Fig.2.} If the thickness of the middle s-electrode, $L_{s}$, is
much larger than the critical thickness of s layer, $L_{sC},$ which separates the different modes of operation in SIsFS structures, the pair
potential $\Delta $ in the s layer is close to that of bulk material.
Note that the critical thickness $L_{sC}$ in an sN (sF) bilayer at a given temperature is generally defined as a minimal thickness
of an s-layer when superconductivity still exists.
In the mode (1) the structure can be considered as a pair of SIs and sFS junctions connected in series.
Therefore, the properties of the structure in parameter range (1) are almost independent on the thickness $L_{s}$
and are determined by the junction with smallest critical current.
It is seen from Fig.\ref{ICRN_LF} that for the given parameter set, $T=0.5T_{C},$ $H=10T_{C},$
$\gamma=1,$ $\gamma_{BI}=1000,$ $\gamma_{BFS}=0.3,$ the critical thickness of the s layer, $L_{sC},$  is close to $3\xi_{S}.$

\emph{Mode (1a) in Fig.\ref{ICRN_LF}.} In the ordinary case of $I_{C\_SIs}\ll I_{C\_sFS}$, the behavior of the
structure coincides with that of conventional SIS junction with one important distinction - the
sFS junction can turn the SIsFS structure into a $\pi $-state. At the same time, other properties
like high $I_{C}R_{N}$ product and sinusoidal current-phase relation are preserved in the $\pi $-state.
Therefore the structure can be called switchable $0$-$\pi $ SIS junction.

\emph{Mode (1b) in Fig.\ref{ICRN_LF}.} Another limiting case is realized
for large $L_{F}$ values and high exchange fields $H$. Namely, the structure transforms into a standard SFS-junction without any influence of tunnel barrier.

\emph{Mode (2) in Fig.\ref{ICRN_LF}.} The absence of superconductivity in the s-electrode in the opposite case (%
$L_{s}\ll L_{sC}$) leads to formation of the complex -InF- weak link area, where n marks the intermediate s film in the normal state.
It results in much smaller critical current value $I_{C}$, with the magnitude close
to that in well-known SIFS junctions \cite{Vasenko}. The dependence of $I_{C}$ on the thickness $L_{s}$ is weak due to large decay length
in the n-region with suppressed superconductivity.

\emph{Mode (3) in Fig.\ref{ICRN_LF}.} Conversely, in the intermediate case ($L_{s}\approx L_{sC}$) the
properties of the structure are extra sensitive to variations of the decay lengths parameters. Within the considered
intermediate thickness range the system may transform from the mode (1) to the mode (2). Moreover, in this situation
the system is sensitive to the F-layer parameters (thickness $L_{F}$ and exchange field $H$),
since these parameters control the suppression of superconductivity in the sF bilayer.

This sensitivity allows one to change an operation mode by changing the parameters such as effective exchange field $H$ and temperature $T.$
Note, that depending on the domain structure of a ferromagnet and morphology of the F-film it might be possible
to control the effective exchange field $H$.

Fig.\ref{Temp1} demonstrates the temperature dependence of the critical current in the
structures with thickness
around critical one (
$L_{sC}=3\xi _{S}$) for different values of
exchange field $H$. These structures are characterized by
the existence of the effective critical temperature $T_{C} ^{*}$ which corresponds to the appearance of
superconductivity in middle s-layer and, correspondingly, to an exponential growth of the current.
Therefore, $T_{C} ^{*}$ may significantly shift during remagnetization of the system
(due to changing of H, as pointed out above). Thus, system can exist either in the superconducting or in the
normal state depending on the history of the application of a magnetic field.
On the other hand, from point of view of practical applications, the $0$-$\pi $ SIS mode (1a) seems more relevant.

Fig.\ref{ICRN_FIG} demonstrates that change of F-layer thickness $L_{F}$
leads to 0-$\pi $ transition. The system can be
switched into a $\pi $-state keeping the value of $I_{C}R_{N}$ product,
i.e., Josephson frequency, on the level characteristic for tunnel SIS junctions. Moreover,
it should be noted that this property of the considered structure is unique.
In the conventional SFS devices in order to reach the $\pi $-state it is necessary to realize
either $L_{F}\gtrsim (2-3) \xi_{F}$ or very large values of the $\gamma_{BFS}\gg 1$ parameter at the SF interfaces.
In both cases the $I_{C}R_{N}$ product in the  $\pi $-state is strongly reduced \cite{RevG,
RevB, RevV, Vasenko}.

\emph{External magnetic field.}
In the parameter range when SIsFS junction is in (1a) mode and far from the $%
0-\pi $ transition, current-phase relation has a sinusoidal form, $I(\varphi
)\approx I_{SIs}\sin \varphi $. To calculate the dependence of $I_{C}$
from external magnetic field, $H_{ext}$, we may use the standard Fraunhofer
expression,
\begin{equation}
I_{C}(H_{ext})=I_{C0}\left\vert \frac{\sin (\pi \Phi /\Phi _{0})}{\pi \Phi /\Phi
_{0}}\right\vert,  \label{Frau0}
\end{equation}
where
\begin{equation}
\Phi =W\left\vert L_{eff}H_{ext}+L_{F}H_{0}N(n_{\uparrow }-n_{\downarrow
})\right\vert
\label{flux}
\end{equation}
is magnetic flux inside of SIsFS junction, $\Phi_{0}$ is flux quantum, $L_{eff}=2\lambda
_{L}+L_{s}+L_{F}+L_{I}$, $\lambda _{L}$ is London penetration depth of S
electrodes, $L_{I}$ is thickness of I layer,
$N$ is the full number of clusters,
$n_{\uparrow, \downarrow} = N_{\uparrow, \downarrow}/N $
are the concentrations of clusters in the F layer oriented parallel ($N_{\uparrow}$) or antiparallel ($N_{\downarrow}$) to the
direction of $H_{ext}$ in the saturation region and $H_{0}$ is an average magnetic field generated by a
single magnetic cluster. In our simple model, $H_{0}$ is assumed to be a constant,
while $n_{\uparrow }$ and $n_{\downarrow }$ are functions of $H_{ext}$.
We assume further that the probability density $p(H_{ext})$ of a flip of a
cluster in $H_{ext}$ is described by a Gaussian distribution
\begin{figure}[tbh]
\begin{center}
\includegraphics[width=8.5cm]{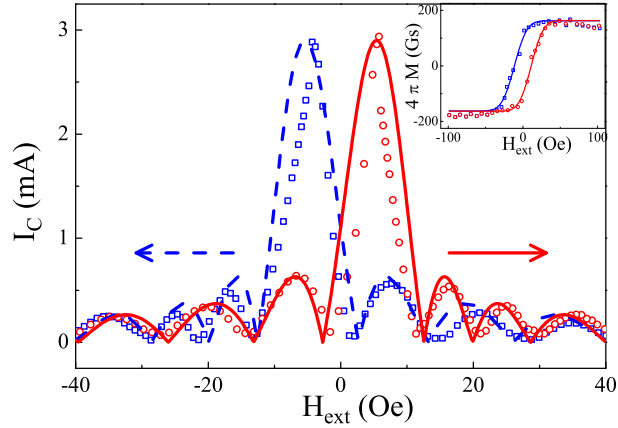}
\end{center}
\caption{Experimental dependence\cite{Vernik} of critical current $I_C$ versus increasing (open circles) and
decreasing (open squares) external magnetic field $H_{ext}$. Solid and dashed lines present the
microscopic fitting of the data\protect. Inset shows
the theoretical and experimental magnetization loops versus external
magnetic field $H_{ext}.$}
\label{Fraun}
\end{figure}
\begin{equation}
p(H_{ext})=(\sqrt{2}/\sqrt{\pi }\delta )\exp \left[ -\left( H_{ext}-H^{\ast
}\right) ^{2}/2\delta ^{2}\right],
\label{Gauss}
\end{equation}
where $H^{\ast }$ is the value of magnetic field, at which the flip of a
cluster takes place. The expectation, $H^{\ast
},$ and the standard deviation, $\delta,$ in (\ref{Gauss}) are independent on $H_{ext}$ values. These parameters,
as well as $H_{0}N$ product (i.e. saturation of the magnetization magnitude) in Eq.(\ref{flux}) can be found by fitting the magnetization curve $4 \pi M(H_{ext}).$

We apply this model to explain  the data experimentally observed in Ref.\cite{Vernik} in SIsFS structures having cross-section
area $10*10$ $\mu$m$^{2}$  and F layer thickness $14$ nm.
Figure 5 demonstrates the experimental dependencies of critical current $I_C$ versus increasing (open circles) and
decreasing (open squares) external magnetic field $H_{ext}$. Solid and dashed lines present the
microscopic fitting of the data. From hysteretic dependencies of F layer magnetization
shown in insert in Fig. \ref{Fraun} we get $H^{\ast
} \approx 11.4$ Oe, $\delta \approx 13$ Oe and  $H_{0}N$ product  $\approx 163$ G.
Remaining fitting parameters can be set from the current magnitude in the main maximum $I_{C0}\approx 2.9$ mA and from the difference between zeros of $I_c(H)$ dependence in the saturation region, where oscillation period depends entirely on the effective length of structure $L_{eff} \approx 150$ nm.

The initial strong
magnetic field $H_{ext}=-\infty $ remagnetize all clusters of F layer into
the homogeneous state $n_{\downarrow }(H_{ext})=1$.
Gradual growth of the
external field provides the conventional Fraunhofer pattern (solid line)
with expected maximum at the positive value $H_{ext}$ corresponding to zero
flux $\Phi =0$. However, the clusters start to flip around the point $%
H_{ext}=H^{*},$
$n_{\uparrow }(H_{ext})=\int\limits_{-\infty }^{H_{ext}}p(H_{ext}^{^{\prime
}})dH_{ext}^{^{\prime }}$.
As a result, the period of Fraunhofer oscillations decreases. Similar situation takes place
during field sweeping in the opposite direction (dashed line), from large positive to negative
values. The densities $n_{\uparrow }(H_{ext})$ and $n_{\downarrow }(H_{ext})=1-n_{\uparrow }(H_{ext})$
can be described by the expression,
$n_{\uparrow }(H_{ext})=0.5\left( 1 + erf\left( (H_{ext}\mp H^{\ast })/\sqrt{%
2}\delta \right) \right),$
for forward and backward remagnitizations, respectively. Here $erf(x)$ is the
error function.

In this paper we  have demonstrated a number of unique properties of SIsFS Josephson junctions. These structures exhibit a large $I_{C}R_{N}$ product in the $\pi$ state comparable to that in SIS tunnel junctions commonly used in SFQ devices.  Moreover, the whole structure behaves as a single junction with respect to an external magnetic field $H_{ext}$. Based on that, we have developed simple model describing the behavior of the critical current in these junctions in external field $H_{ext}$ taking into account remagnetization of the F-layer. The model explains asymmetric Fraunhofer oscillations $I_{C}(H_{ext})$ in Nb-Al/AlO${_x}$-Nb-Pd$_{0.99}$Fe$_{0.01}$-Nb junctions reported in \cite{Larkin, Vernik, Ryazanov3}. These effects provide the possibility to realize magnetic memory compatible with energy-efficient SFQ digital circuits \cite{Mukhanov1} with high switching speed.

This work was supported by the Russian Foundation for Basic Research, Russian Ministry of Education and Science,
Dynasty Foundation, Scholarship of the President of the Russian Federation and Dutch FOM.

\end{document}